\documentclass{raa01}
\usepackage{graphicx, times}  
\usepackage{amsmath}   
\usepackage{natbib}
\usepackage{amssymb,amsmath}
\bibpunct{(}{)}{;}{a}{}{,}

\begin{document}

   \title{Cosmic microwave background constraints on the tensor-to-scalar ratio}

\volnopage{{\bf 2014} Vol.\ {\bf 14} No. {\bf 6}, 635--647~~
 {\small doi: 10.1088/1674--4527/14/6/003}}
   \setcounter{page}{635}

   \author{King Lau
        \and Jia-Yu Tang
        \and Ming-Chung Chu
         }

   \institute{Department of Physics and Institute of Theoretical Physics,
   The Chinese University of Hong Kong, Hong Kong, China; {\it klau@phy.cuhk.edu.hk}\\
\vs\no
   {\small Received 2013 October 22; accepted 2013 December 27}}

\abstract{  One of the main goals of modern {c}osmic {m}icrowave
{b}ackground (CMB) missions is to measure the tensor-to-scalar
ratio $r$ accurately to constrain inflation models. Due to
ignorance {about} the reionization history $X_{e}(z)$, this
analysis is usually done by assuming an instantaneous reionization
$X_{e}(z)$ which{,} however{,} can bias the best-fit value of $r$.
Moreover, due to the strong mixing of B-mode and E-mode
polarizations in cut-sky measurements, multiplying the sky
coverage fraction $f_{\textup{sky}}$ {by} the full-sky likelihood
would not give satisfactory results. In this work, we forecast
constraints on $r$ for the \textit{Planck} mission taking into
account the general reionization scenario and cut-sky effects. Our
results show that by applying an N-point interpolation analysis
{to} the reionization history, the bias induced by the{ assumption
of} instantaneous reionization is removed and the value of $r$ is
constrained within $5\%$ error level, if the true value of $r$ is
greater than about 0.1.  }

   \authorrunning{K. L{au}, J. T{ang} \& M. C. C{hu}} 
   \titlerunning{
   Cosmic Microwave Background Constraints on
   {the }Tensor-to-scalar Ratio}  

\maketitle

\section{Introduction}
\label{Introduction}

Inflation (Guth~\citeyear{inflation1};
Linde~\citeyear{inflation2}; Albrecht \&
Steinhardt~\citeyear{inflation3}) is now the leading paradigm in
cosmology. The inflation scenarios {have been} proposed to solve
the{ problems of} horizon, flatness{ and} magnetic monopoles and
explain the generation of primordial perturbations in {the }early
Universe. Most inflation models predict two types of initial
perturbations: scalar and tensor. The scalar perturbations are
adiabatic, nearly Gaussian and close to being scale-invariant,
which are consistent with a series of observations (Hu \&
White~\citeyear{predbyinflation1}; Spergel \&
Zaldarriaga~\citeyear{predbyinflation2}; Hu et
al.~\citeyear{predbyinflation3}; Peiris et
al.~\citeyear{predbyinflation4}; Spergel et
al.~\citeyear{predbyinflation5}; Hinshaw et
al.~\citeyear{WMAP9_1}). The tensor perturbations produced during
inflation, also known as gravitational waves, can be quantified as
a tensor-to-scalar ratio $r$. Therefore, a non-zero $r$ is
considered {to be} important evidence of inflation if {it is
}observed. Since tensor perturbations can be detected in{ a}
large-scale temperature power spectrum and should have left an
imprint on the B-mode polarization of the {c}osmic {m}icrowave
{b}ackground (CMB) (Seljak \& Zaldarriaga~\citeyear{rinCMB1};
Kamionkowski et al.~\citeyear{rinCMB2}), constraining $r$ is one
of the main goals of modern CMB surveys. Recent data from the
{nine} year result of{ the} Wilkinson Microwave Anisotropy Probe
(WMAP9) and South Pole Telescope give the latest constraints of
$r<0.13$ and $r<0.11$ at{ the} 95\% confidence level (CL)
respectively without{ a measurement of the} B-mode polarization
(Story et al.~\citeyear{SPT}; Hinshaw et al.~\citeyear{WMAP9_1};
Bennett et al.~\citeyear{WMAP9_2}). Although \textit{Planck}'s
results were released in March 2013 (\citealt{PLANCK1}), its
polarization data, which {are} crucial for constraints on $r$,
{are} not yet available. Their current approach combines
\textit{Planck}'s{ measurements of} temperature anisotropy with
the WMAP large-angle polarization to constrain inflation, giving
an upper limit{ of} $r<0.11$ {as a} 95\% CL
(\citealt{PLANCK_inflation}).

These constraints for $r$ were obtained by assuming an
instantaneous model for reionization history, but $X_{e}(z)$, the
average ionized fraction at redshift $z$, is rather uncertain.
Various sources such as star formation (Springel \&
Hernquist~\citeyear{starformation1}; Bunker et
al.~\citeyear{starformation2}), massive black holes (Sasaki \&
Umemura~\citeyear{massiveBH1}) and dark matter decay (Mapelli et
al.~\citeyear{darkmatterdecay2}; Belikov \&
Hooper~\citeyear{darkmatterdecay1}) {have been} suggested to
provide the energy flux{ necessary for} reionizing hydrogen. CMB
and {q}uasar observations show that recombination occurs at
redshift $z_{*} \sim 1100$ and the Universe must {have been} fully
reionized at $z \sim 6$ (Becker et al.~\citeyear{reion1}; Fan et
al.~\citeyear{reion2}), but there is no detailed knowledge about
the evolution of $X_{e}(z)$ between these two {eras}. The
constraint imposed on $X_{e}(z)$ by current CMB measurement{s} is
also poor. The CMB temperature power spectrum $C_{l}^{\rm TT}$
{only }gives a strong constraint on $A_{\rm s}e^{-2\tau}$, where
$A_{\rm s}$ and $\tau$ are the scalar amplitude and optical depth
respectively. Even if we can break the degeneracy between these
{two} parameters, we can only get the information {from} $\tau$,
where
\begin{equation}
\tau \propto \int^{z_{*}}_{0} \left( X_e(z) \sqrt{1+z} \right)
dz\, , \label{cal_tau}
\end{equation}
but not $X_e(z)$ itself. The CMB E-mode polarization spectrum
$C_{l}^{EE}$ {only }has a weak dependence on the reionization history,
and thus {an} attempt {at} constraining $X_{e}(z)$ by {the }current
E-mode polarization measurement does not give any satisfactory
result (Lewis et al.~\citeyear{EEconstrainXe}).

Several studies have considered the effects of uncertainties in
$X_{e}(z)$ on cosmological parameter estimation. To
paramet{e}rize the reionization history,  Lewis et al. assume{ a}
constant ionization fraction in finite redshift bins and join the
bins using a tanh function (Lewis et al.~\citeyear{EEconstrainXe}).
Mortonson et al. propose that the reionization history can be
expressed as a linear combination of finite numbers of principal
components $S_{\mu}$ extracted from the Fisher{ information} {m}atrix that describes
the dependence of E-mode polarization on reionization, so that the
amplitudes of $S_{\mu}$ are parameters for $X_{e}$ (Mortonson \&
Hu~\citeyear{Humodelindependent}). To consider the general
reionization scenario, Pandolfi et al. apply these two methods in
their analysis to constrain the inflation parameters by WMAP7 data
(temperature and E-mode polarization only) (Pandolfi et
al.~\citeyear{Impact}) while the PLANCK Collaboration
(\citealt{PLANCK_inflation}) {only }adopts the method by Mortonson et al. A recent study investigates how instantaneous-like
reionization models affect the estimation of{ all} cosmological parameters
from \textit{Planck}-quality CMB data except{ for} $r$
(\citealt{Kinney2012}). To account for the fact that parts of the
sky are masked to eliminate foreground contaminations, they multiply
the sky coverage fraction $f_{\textup{sky}}$
($f_{\textup{sky}}=0.65$ for \textit{Planck}) {by} their full-sky
likelihood.

In this paper, we explore how well $r$ can be constrained by the
\textit{Planck} mission with a general parametrization of the
reionization history. In Section~\ref{Degeneracy}, we discuss the
degeneracy between the reionization history $X_{e}(z)$ and $r$ in
the full-sky CMB power spectra and the necessity of using both
temperature and polarization power spectra for constraining $r$.
Then we make a full-sky forecast and conclude that a bias is
possibly introduced in $r$ if an incorrect reionization assumption
is applied in the Markov Chain Monte Carlo (MCMC) analysis
(Kosowsky et al.~\citeyear{montecarlo}). The N-point linear
interpolation method for reionization is also introduced. We then
discuss the significance of strong mixing of E-mode and B-mode
polarizations and show that the simple $f_{\textup{sky}}$
modification is unrealistic in constraining $r$ using cut-sky
power spectra in Section~\ref{Likelihood}. Thus we apply the
Hamimeche and Lewis likelihood approximation which can handle the
CMB temperature-polarization correlation for high $l$'s in{ the}
cut-sky. In Section~\ref{Forecast}, we present the forecast of
\textit{Planck}'s constraint on $r$ using a general reionization
representation.

\newpage

\section{Degeneracy among reionization history, $\lowercase{n}_{\rm T}$,
\lowercase{$r$} and \lowercase{$n_{\rm s}$}} \label{Degeneracy}

In this study, we consider single-field inflation with the
slow-roll approximation. Conventionally, the power spectra of
scalar perturbations $P_{\mathcal{R}}$ and tensor fluctuations
$P_{h}$ have the functional form
\begin{eqnarray}
k^{3}P_{\mathcal{R}}(k)&\propto& k^{n_{\rm s}-1}\, , \\
k^{3}P_{h}(k) &\propto& k^{n_{\rm T}},
\end{eqnarray}
in which $n_{\rm s}$ and $n_{\rm T}$ are {the }spectral index and
 tensor tilt, respectively. Hence, the tensor-to-scalar ratio $r$ is defined as
\begin{equation}
r\equiv\frac{P_{h}(k_{0})}{P_{\mathcal{R}}(k_{0})},
\end{equation}
where $k_{0}$ is the pivot scale. Our choice for it is $k_{0}
=0.002 $~Mpc$^{-1}$. Moreover, in the simple slow-roll inflation
model, there is a well-known consistency relation
(Kinney~\citeyear{inflationdisc.2})
\begin{equation}\label{consistency relation}
n_{\rm T} = -\frac{r}{8},
\end{equation}
which is correct at first order in the slow-roll parameters. These
parameters are of interest {for studying the} CMB {because they
give} an accurate measurement of $r$ and spectral index $n_{\rm
s}$ can discriminate among inflation models (Dodelson et
al.~\citeyear{inflationdisc.1};
Kinney~\citeyear{inflationdisc.2}).

\begin{figure}

\vs \centering
\includegraphics[width=6.5cm]{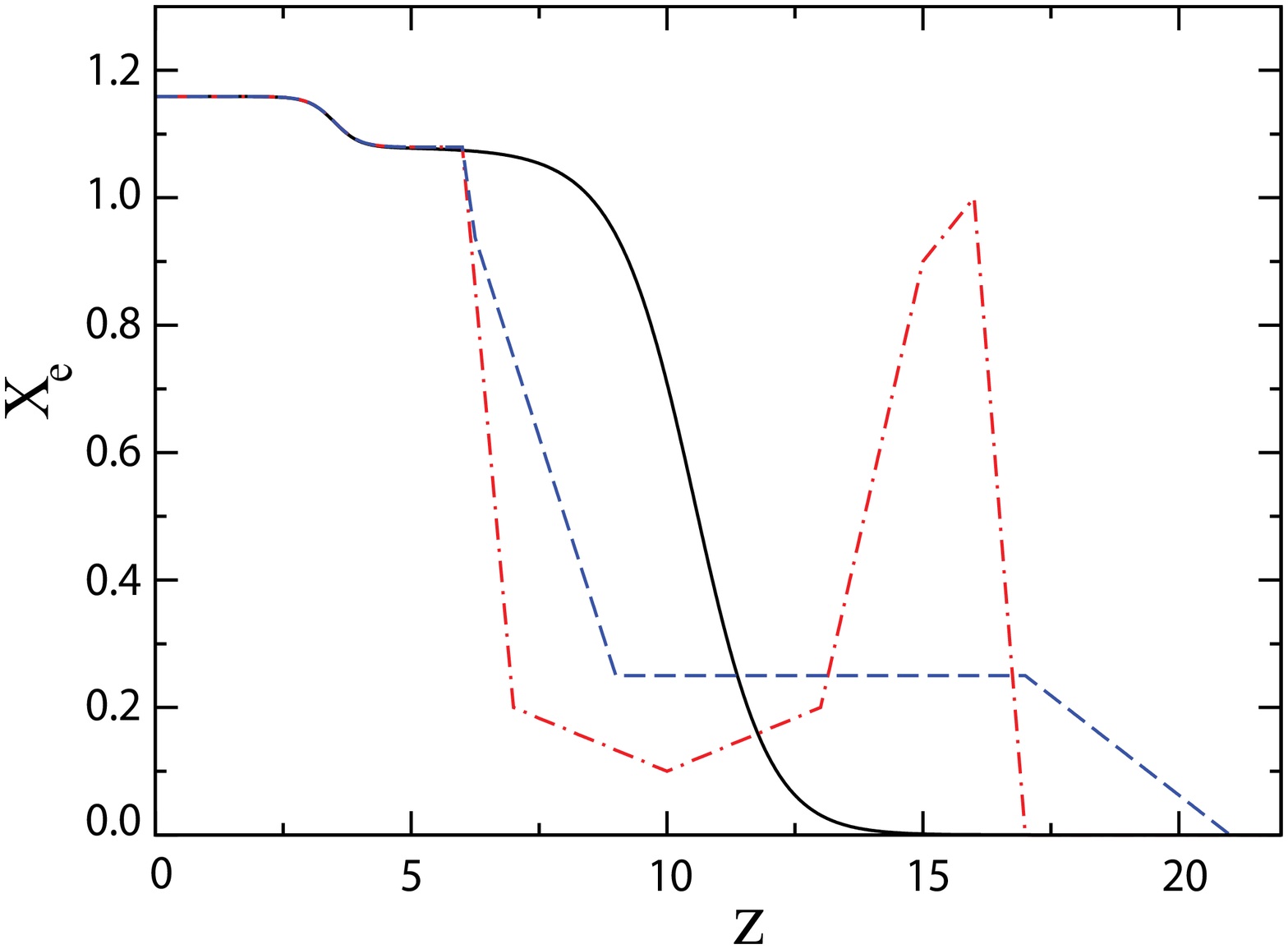}

\caption{\baselineskip 3.6mm The {three} fiducial reionization
models considered in this paper: instantaneous reionization ({\it
solid line}), two-step reionization ({\it dashed line}) and double
reionization ({\it dashed-dotted line}). All of them are assumed
to reach full H ionization at $z \sim 6$ and have late time He
reionization at $z \sim 3$. All of them give $\tau=0.089$.}
\label{Xe_detail_compare3}

\vs \centering
\includegraphics[width=6.cm]{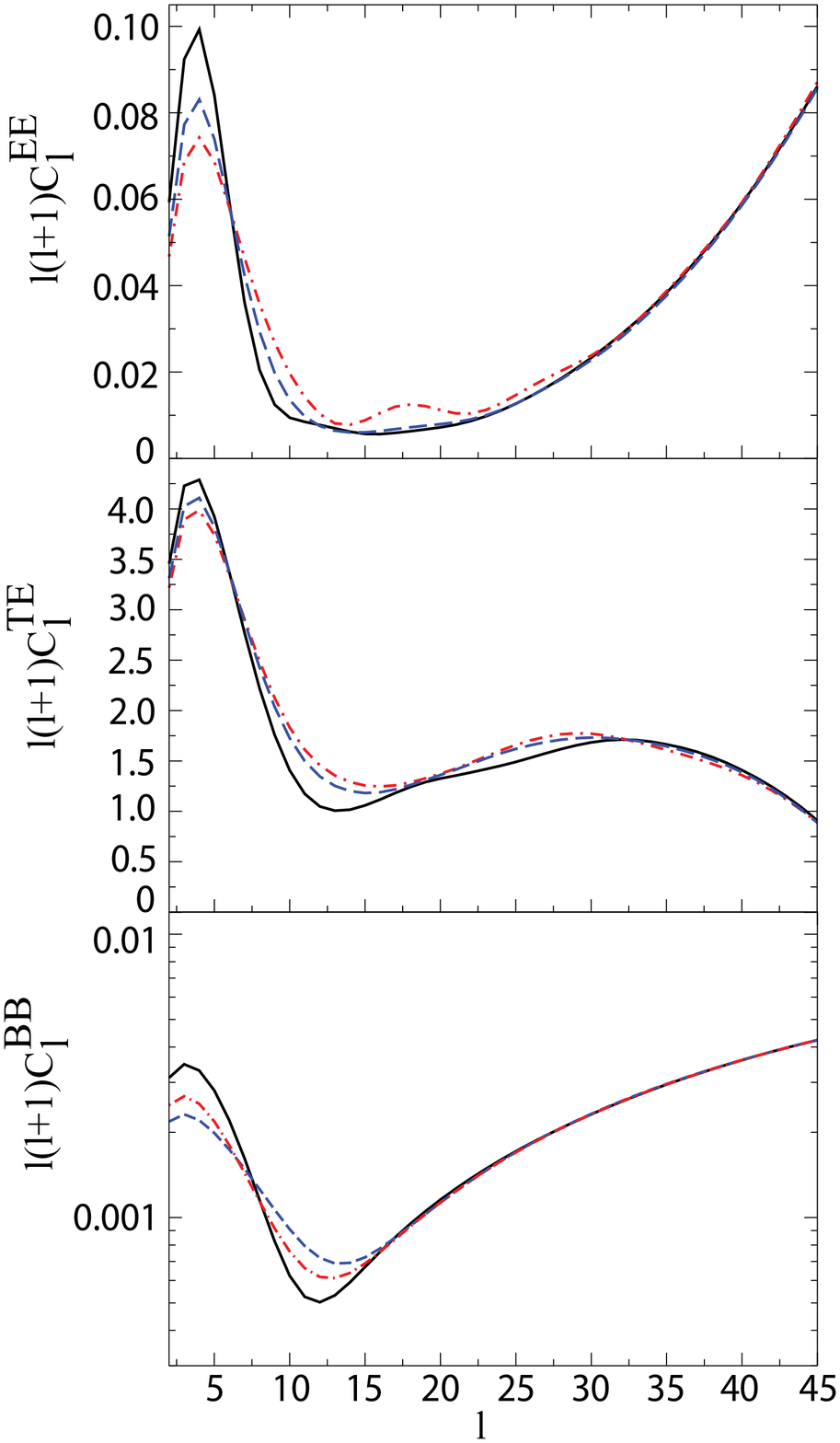}

\caption{\baselineskip 3.6mm The CMB E-mode polarization power
spectrum, temperature-E cross spectrum and B-mode polarization power
spectrum (all generated by standard $\Lambda$CDM best-fit WMAP9
parameters with $r$=0.1) for three reionization histories:
instantaneous ({\it solid line}), two-step ({\it dashed line}) and
double reionization ({\it dash{ed}-dotted line}) as shown in
Fig.~\ref{Xe_detail_compare3}. They are sensitive to the
reionization history for $l \leq 30$.} \label{EE TE
BB depends on Xe}
\end{figure}

Previous analysis of CMB data usually assumes the reionization
history to be instantaneous,
\begin{equation}
X_e(z) \propto
\frac{1}{2}+\frac{1}{2}\textup{tanh}\left(\frac{z_{\rm
re}-z}{\Delta_{z}}\right), \label{instant_re}
\end{equation}
where $\Delta_z$ is a width parameter and $z_{\rm re}$ is the
redshift at which reionization occurs. Here, we take
$\Delta_z=1.5$. Combining Equation~(\ref{instant_re}) with
Equation~(\ref{cal_tau}), we can {utilize} CMB data to make{ an}
inference for the parameters $z_{\rm re}$ and $\tau$. However, it
has been pointed out that $C_{l}^{\rm EE}$ and $C_{l}^{\rm TE}$,
unlike $C_{l}^{\rm TT}$, depend not only on $\tau$ but also on the
detailed evolution of $X_{e}(z)$ (Lewis et
al.~\citeyear{EEconstrainXe}), especially for $l<30$. To compare
the impacts of the reionization history on CMB polarization power
spectra, we consider the instantaneous model and two other
physically acceptable reionization models, double reionization and
two-step reionization, as illustrated in
Figure~\ref{Xe_detail_compare3}. 
 The former model {has}
two instantaneous reionization{s} occur{ring} at $z=7, 17$ and a
sudden{, midway} recombination{,} while the latter describes a
reionization process with a long{,} intermediate pause. All of
them give $\tau=0.089$ and are assumed to reach full H ionization
at $z \sim 6$, as well as a late time He reionization at $z \sim
3$. The corresponding $C_{l}^{\rm EE}$, $C_{l}^{\rm TE}$ and
$C_{l}^{\rm BB}$ are shown in Figure~\ref{EE TE BB depends on Xe}.
 The distinct differences {in the} three curves for
$l<30$ indicate the dependences of CMB polarization power spectra
on the reionization history $X_{e}(z)$. Although double
reionization is rather disfavored by current observation{s} (Zahn
et al.~\citeyear{doublepeak}), it helps to demonstrate the bias on
$r$ if a{n} {incorrect} reionization model is used.

To examine{ the} possible bias of parameters by a{n} {incorrect} reionization
model, we run a simple test on two sets of full-sky CMB power
spectra {that have quality comparable to} \textit{Planck}. These two sets of power
spectra are generated using the standard WMAP9 best-fit cosmological
parameters but with the two-step reionization and double
reionization model respectively. Meanwhile, we still use an
instantaneous reionization model and perform MCMC analysis to
estimate $\tau$ from these power spectra (refer to
Sect.~\ref{Forecast} for details about MCMC fitting).

Figure~\ref{bias in tau and zre fullsky} 
 shows the
probability functions of $\tau$ and $z_{\rm re}$ after
marginalizing over all other parameters. The dashed and
dashed-dotted lines indicate the results {using} two-step
reionization and double reionization, respectively. Both
probability functions show a sharp convergence, but a bias in
$\tau$ is introduced relative to the fiducial value of the optical
depth $\tau = 0.089$, indicated by the vertical line. Moreover,
the bias with the double reionization is larger, which reflects
the larger difference between the instantaneous model and the
double reionization model. As there is a $\tau-n_{\rm T}$
degeneracy on the B-mode polarization spectrum (Mortonson \&
Hu~\citeyear{Hu2008}), if $\tau$ is biased to a larger value, the
corresponding $n_{\rm T}$ becomes less negative; because $r$ and
$n_{\rm T}$ are correlated in CMB constraints, a bias in the
estimation of $r$ is introduced.

\begin{figure}
\centering
\includegraphics[width=3.6cm]
{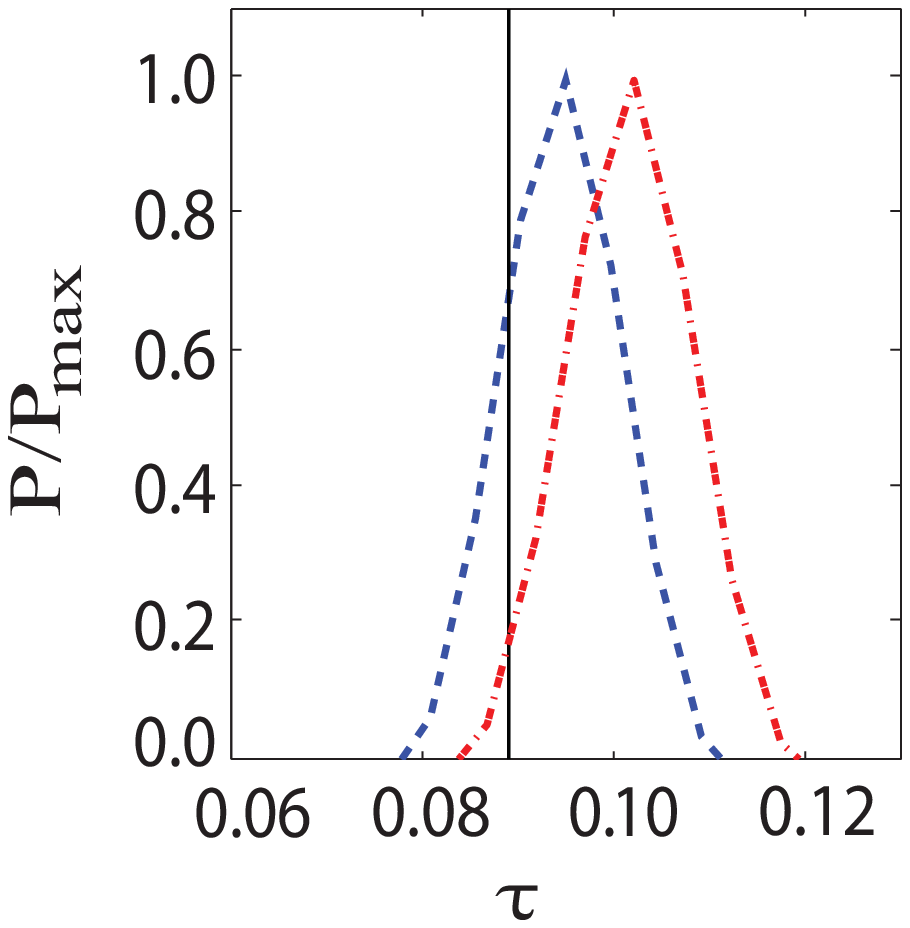}~~~~~~~~
\includegraphics[width=3.6cm]{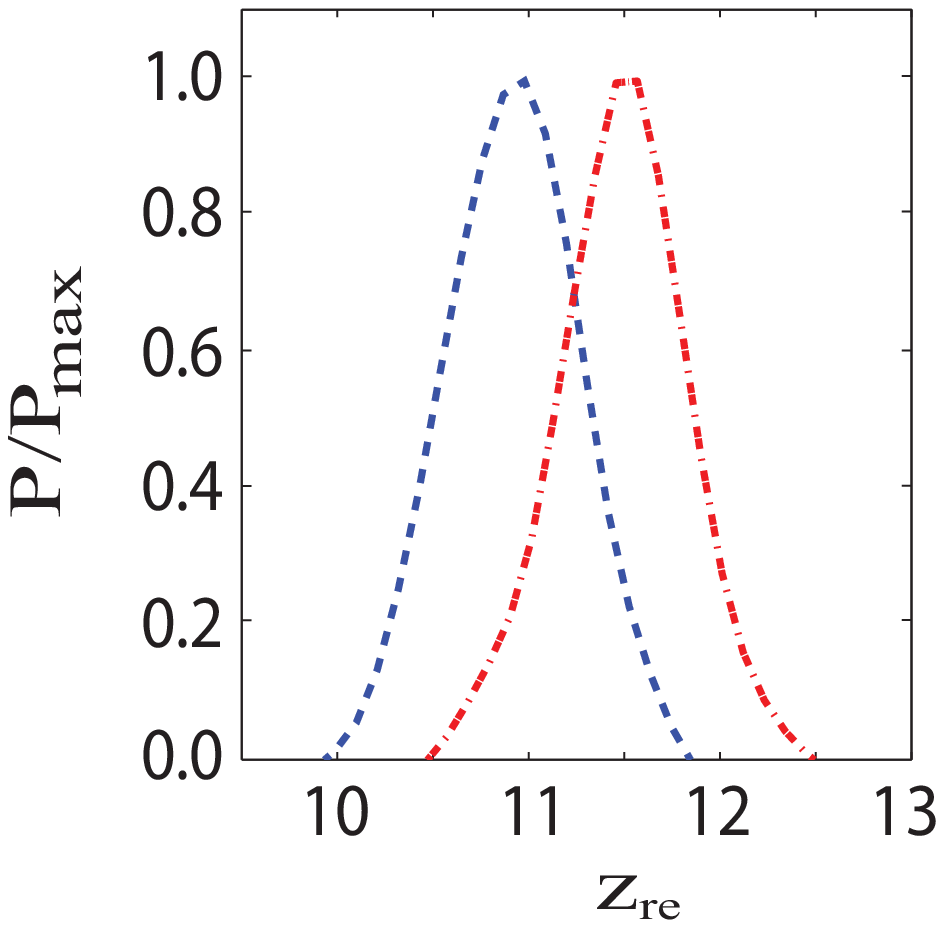}

\vspace{-3mm} \caption{\baselineskip 3.6mm Probability functions
$P$ (relative to the peak value $P_{\max}$)
of the MCMC fitting results for the optical depth $\tau$ ({\it
left}) and reionization redshift $z_{\rm re}$ ({\it right}) from
two sets of expected \textit{Planck} full-sky CMB power spectra
(including temperature and polarization). We calculate these power
spectra using the two-step reionization ({\it dashed line}) and
double reionization ({\it dashed-dotted line}), while assum{ing}
the instantaneous reionization history to make{ the} MCMC fitting.
Both probability functions show a sharp convergence but a bias is
introduced relative to the fiducial value of the optical depth
($\tau = 0.089$) indicated by the vertical line.} \label{bias in
tau and zre fullsky}
\end{figure}

Therefore, to avoid any assumption on reionization history, we
modify an N-point parametrization of reionization which {was}
initially proposed by Lewis et al. (\citeyear{EEconstrainXe}). We
fix $X_{e}(z=6)=1.08$ (with He reionization) and $X_{e}(z=22)=0$,
and we insert $N$ floating point values $\{ {X_{e}(z_{i})}
\}_{i=1}^{N}$ at redshift 
\begin{equation}
z_{i}=6+i(22-6)/(N+1), \;\;\;\;\;\; i=1, \cdots, N.\label{Npoint}
\end{equation}
Then the whole reionization history in $6 \leq z \leq 22$ is the
linear interpolation among these points. This method introduces $N$
reionization points, which are extra parameters in the MCMC analysis
together with the cosmological parameters, and they are free to vary
in{ the range} [0,1] (the physical range of $X_e(z)$). We assume a late-time
He reionization{ occurred} at $z \sim 3$.

\section{Likelihood of cut-sky CMB}
\label{Likelihood}

To make {a }forecast from fiducial CMB power spectra, we perform a
TT-TE-EE-BB joint analysis covering both small and large scales
with the N-point method. It is important to include the
$C_{l}^{\rm TT}$ power spectrum to constrain the baryonic density
$\Omega_{\rm b}$, dark matter density $\Omega_c$, Hubble parameter
$H$ and $A_{\rm s}e^{-2\tau}$ well. As tensor perturbations
contribute to the CMB temperature power spectrum at large scales,
its effect is degenerate with the change in $n_{\rm s}$ in
large-scale CMB measurements. Therefore, the $C_{l}^{\rm TT}$
power spectrum {at} small scales can help break this degeneracy by
constraining $n_{\rm s}$ well. It is also necessary to include
$C_{l}^{\rm EE}$ and $C_{l}^{\rm BB}$ power spectra {at} large
scales, and drop the assumption of instantaneous reionization to
break the $\tau-r$ degeneracy as we discussed in Section
\ref{Degeneracy}. Although $C_{l}^{\rm TE}$ has a weak dependence
on reionization history, as it is a cross spectrum, its noise is
much reduced, and thus it helps to break this degeneracy in a
joint analysis.

Assuming that the CMB field is Gaussian and isotropic, the full-sky
maximum likelihood $\mathcal{L}$ for the measured TEB correlation
spectra $\hat{C}_{l}$ is
\begin{equation}
-2\textup{ln}\mathcal{L}=\sum_{l_{\textup{min}}}^{l_{\textup{max}}} (2l+1)[\textup{Tr}(\mathbf{\hat{C}}_{l}\mathbf{C}_{l}^{-1})-\textup{ln}|\mathbf{\hat{C}}_{l}\mathbf{C}_{l}^{-1}|-3],
\label{likelihood}
\end{equation}
where
\begin{equation}
\mathbf{C}_{l}=
\begin{pmatrix}
C_{l}^{\rm TT} & C_{l}^{\rm TE} & C_{l}^{\rm TB} \\[0.3em]
C_{l}^{\rm TE} & C_{l}^{\rm EE} & C_{l}^{\rm EB} \\[0.3em]
C_{l}^{\rm TB} & C_{l}^{\rm EB} & C_{l}^{\rm BB}
\end{pmatrix}
\end{equation}
is the matrix of theoretical power spectra and
$\mathbf{\hat{C}}_{l}$ is defined similarly.

In practice, masks are applied for both temperature and polarization data (Jarosik et al.~\citeyear{WMAPappliedmasks}), to exclude the part of the sky map contaminated by the astrophysical foreground, mainly the Galactic plane. To have a cut-sky forecast, the sky coverage fraction  $f_{\textup{sky}}$ is usually included as a factor in Equation~(\ref{likelihood}) to account for the fractional loss of $\hat{C}_{l}$ power spectra in {the }cut-sky. While $f_{\textup{sky}}$ modification works well for the temperature power spectrum, it is not sufficient for the polarization spectra, but the ${C}_{l}^{BB}$ power spectrum is crucial for the constraint of $r$. Burigana et al. have considered a toy model to include the impact of the foreground contamination in full-sky likelihood analysis (Burigana et al.~\citeyear{burigana}). This method, however, will introduce uncertainties in modeling the noise residuals and therefore 
$r$ and $X_e$. With this consideration, in this work, we use the pseudo-$C_{l}$ as the estimators and apply the likelihood approximation introduced by Hamimeche and Lewis (hereafter H.L. likelihood) (Hamimeche \& Lewis~\citeyear{Lewis2008}). The pseudo power spectrum $P(C_{l})$, defined as the power spectrum over a masked sky map, is related to the full-sky CMB spectrum by the relation (Kogut et al.~\citeyear{kogutmaskedmatrix})
\begin{equation}
\left\langle\begin{pmatrix}
P(C_{l}^{\rm TT}) \\[0.3em]
P(C_{l}^{\rm TE}) \\[0.3em]
P(C_{l}^{\rm EE}) \\[0.3em]
P(C_{l}^{\rm BB})
\end{pmatrix}\right\rangle=
\sum_{l^{'}}
\begin{pmatrix}
M_{ll^{'}}^{\rm TT} & 0 & 0 & 0\\[0.3em]
0 & M_{ll^{'}}^{\rm TE} & 0 & 0\\[0.3em]
0 & 0 & M_{ll^{'}}^{\rm EE} & M_{ll^{'}}^{\rm EB}\\[0.3em]
0 & 0 & M_{ll^{'}}^{\rm BE} & M_{ll^{'}}^{\rm BB}
\end{pmatrix}
\begin{pmatrix}
C_{l^{'}}^{\rm TT} \\[0.3em]
C_{l^{'}}^{\rm TE} \\[0.3em]
C_{l^{'}}^{\rm EE} \\[0.3em]
C_{l^{'}}^{\rm BB}
\end{pmatrix},
\end{equation}
where $\langle...\rangle$ denotes the expectation value and $\{
M^{XY}_{ll^{'}} \}$ are the coupling mask matrices. Details about
this likelihood {are} described in{ the} {Appendix}.

\begin{figure}

\vs \centering
\includegraphics[width=4.3cm]
{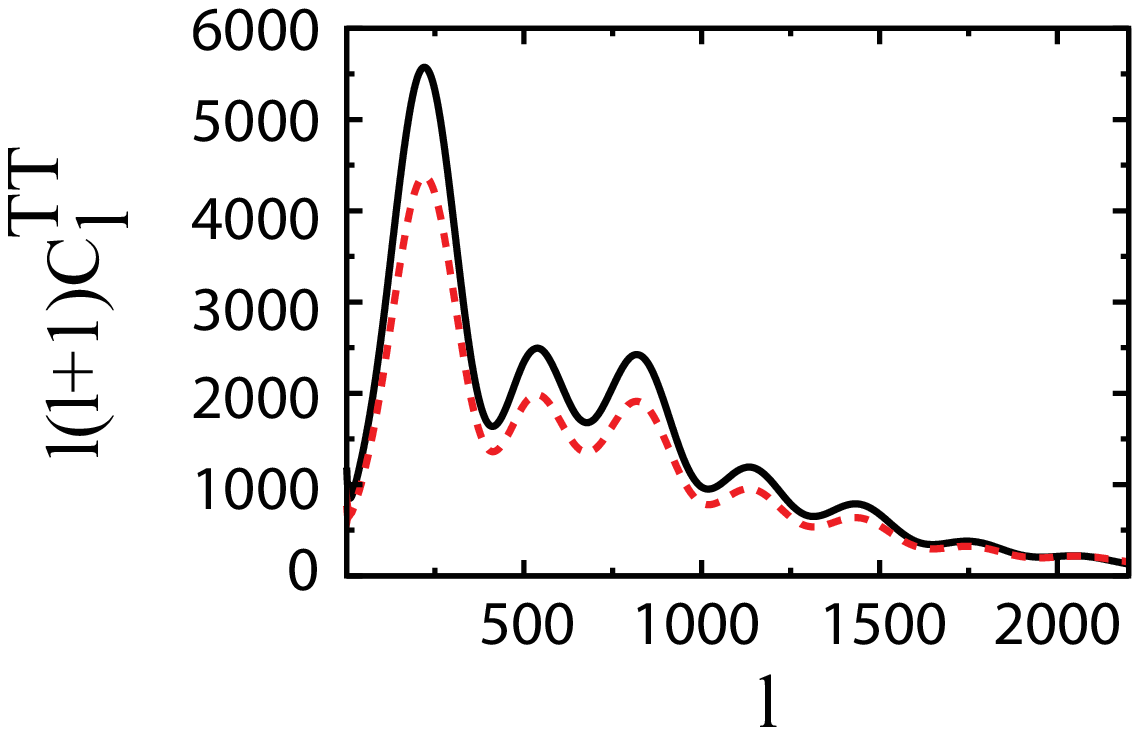} \qquad\qquad
\includegraphics[width=4.3cm]{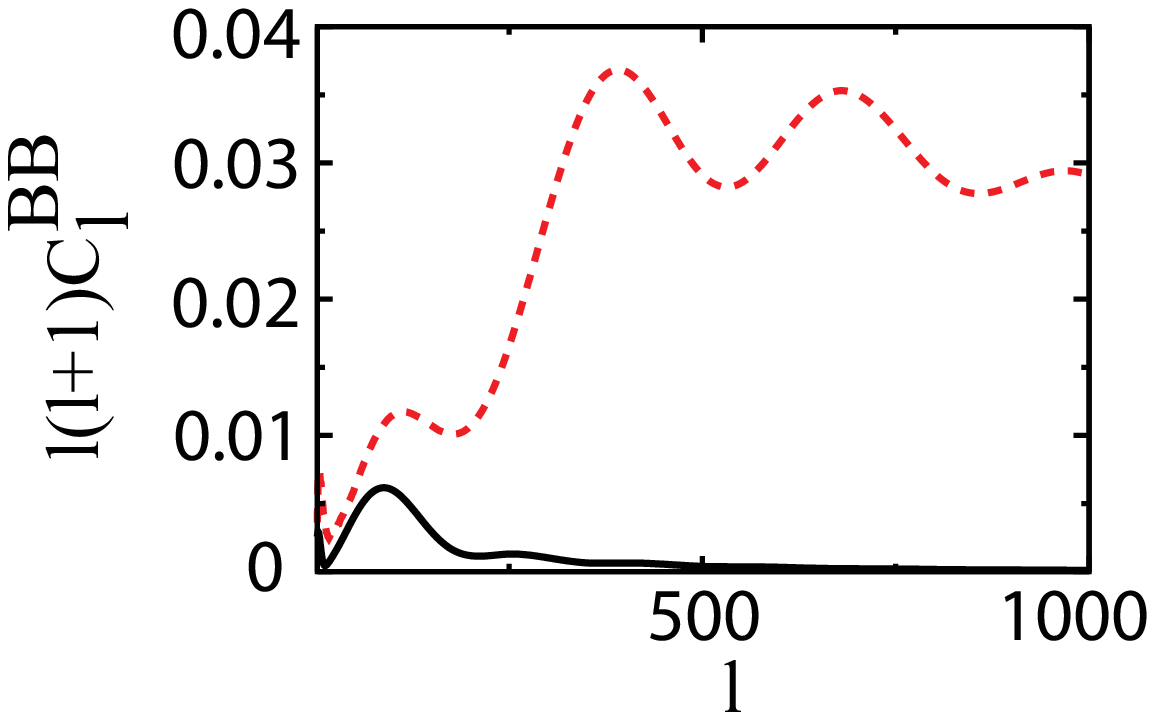}

\vspace{-3mm} \caption{\baselineskip 3.6mm The CMB temperature
({\it left}) and B-mode polarization power spectrum ({\it right})
for full-sky ({\it solid lines}) and their pseudo power spectra
for cut-sky ({\it dashed lines}), for the fiducial model of
$r=0.1$ and standard cosmological parameters. $P(C_{l}^{\rm BB})$
is increased by the mask due to mixing with $C_{l}^{\rm EE}$,
making the simple $f_{\textup{sky}}$ modification unrealistic. The
masks applied by WMAP are used in this illustration.}
\label{masked graph}
\end{figure}

Before we proceed, we illustrate in Figure~\ref{masked graph}
 why a simple $f_{\textup{sky}}$ modification of the
full-sky likelihood would not give a realistic constraint on $r$.
Figure~$\ref{masked graph}$ compares the full-sky $C_{l}^{\rm TT}$
and $C_{l}^{\rm BB}$ (solid lines) and their corresponding pseudo
power spectrum in {the }cut-sky (dashed lines), for $r=0.1$. We
apply the same masks released by the WMAP team\footnote{The masks
for WMAP are available at {\it 
http://lambda.gsfc.nasa.gov/product/map/dr4/masks\_get.cfm}},
which {are} equivalent to $f_{\rm sky}=0.65$. As expected,
$P(C_{l}^{\rm TT})$ is approximately reduced by a factor of
$f_{\rm sky}$ compared with $C_{l}^{\rm TT}$. However,
$P(C_{l}^{\rm BB})$ is significantly increased, due to the mixing
between $C_{l}^{\rm EE}$ and $C_{l}^{\rm BB}$ and the coupling
among different $l$-modes.

\section{Forecast for \textit{Planck}}
\label{Forecast}

To forecast the constraints {on} $r$ by the \textit{Planck} survey
with a general reionization scenario, we perform MCMC analysis with
the H.L. likelihood discussed in{ the} {Appendix}. The expected
\textit{Planck} pseudo power spectra $\hat{C_l}$ from the fiducial
power spectra $C_{l}^{\textup{fid}}$ are calculated as
\begin{equation}
\label{expected_spectrum}
\hat{C}_{l}^{XY}= P(b_{l}^{2}C_{l}^{\textup{fid}XY}+N_{l}^{XY})
\end{equation}
for $X$, $Y$ = $T$, $E$ or $B$, where
\begin{equation}
b_{l}^{2}=\textup{exp}\left(\frac{-l(l+1)(\theta_{\textup{fwhm}}/\textup{rad})^{2}}{8\textup{ln}2}\right),
\end{equation}
the beam width $\theta_{\textup{fwhm}}=7.1'$ and the noises
$N_{l}^{XY}$ for{ the} $C_{l}^{XY}$ spectrum are $N_{l}^{\rm
TT}=1.5\times 10^{-4} \mu{\textup{K}}^2$,
$N_{l}^{EE}=N_{l}^{BB}=3.61N_{l}^{\rm TT}$ (PLANCK
Collaboration~\citeyear{Planckveryearly}) and $N_{l}^{TE}=0$. We
use the frequency band cent{e}red at 143 GHz and assume that there
is no correlation among the random noise fields.
$C_{l}^{\textup{fid}}$ is computed using CAMB\footnote{CAMB is
available at {\it http://camb.info/}} (Lewis et
al.~\citeyear{CAMB1}). The fiducial model used to compute
$C_{l}^{\textup{fid}}$ is $\Omega_{b}h^2=0.0227,
\Omega_{c}h^2=0.108, n_{\rm s}=0.961, 100\theta=1.040137${ and
 }$A_{\rm s}=2.41\times10^{-9}$, using standard notations for the
cosmological parameters. Two fiducial models of reionization, the
double reionization and two-step reionization{,}  are considered
in our analysis, which are shown in
Figure~\ref{Xe_detail_compare3}.  In addition, we investigate
three cases {for} $r$: $r=0.05,0.1${ and }$0.15$ and their
corresponding tensor tilt is taken as $n_{T}=-r/8$. Thus, {six}
sets of pseudo power spectra $\hat{C}_{l}^{XY}$ are generated.

We perform the MCMC analysis using the modified version of
CosmoMC\footnote{CosmoMC is available at {\it
http://cosmologist.info/cosmomc/}} (Lewis \&
Bridle~\citeyear{COSMOMC1}) by Mortonson and Hu\footnote{Mortonson
and Hu's program is available at {\it
http://background.uchicago.edu/camb\_rpc/}} (Mortonson \&
Hu~\citeyear{Hu2008}). To study the impact of reionization history
on estimation as discussed in Section~\ref{Degeneracy}, two
treatments on $X_e$ are applied. For the first, we use the
instantaneous reionization modeled as in
Equation~(\ref{instant_re}). The varied parameters are $\{
\Omega_{b}h^2, \Omega_{c}h^2, \theta, n_{\rm s}, n_{T},
\textup{log}(10^{10}A_{\rm s}), r,  \tau\}$. The second treatment
uses the N-point method defined in Equation~(\ref{Npoint}); we
further modify the CosmoMC program to vary parameters
$\{\Omega_{b}h^2, \Omega_{c}h^2, \theta, n_{\rm s}, n_{T},
\textup{log}(10^{10}A_{\rm s}), r, \{ {X_{e}(z_{i})}
\}_{i=1}^{N}\}$ for $N=7$. The likelihood is summed over
$l_{\textup{min}}=2$ and $l_{\textup{max}}=2200$. We are aware
that the H.L. likelihood is less reliable for pseudo-$C_{l}$ at
low-$l$ range, and Hamimeche \& Lewis (\citeyear{Lewis2008}) state
that exact likelihood calculation is feasible at low-$l$ when
realistic foreground contamination is carefully considered.
However, this investigation is beyond the scope of this paper. In
addition, the CMB polarization power spectra at low-$l$ are
essential for breaking the degeneracy between the reionization
history and inflation parameter $r$. Our results show that if
$l_{\textup{min}}$ is taken as $10$, only $\tau$ is biased by
$\sim 1\sigma$ by the choice of $l_\textup{min}$ (smaller
$l_\textup{min}$ gives{ a} better constraint), while similar
best-fit values of $r$ and other cosmological parameters with
slightly larger uncertainties are obtained. Therefore, we extend
the application of H.L. likelihood to $l_{\textup{min}}=2$ and
focus on the results based on this condition.

Figure~\ref{Npt 01 015 reion1 to Convention}  and Figure~\ref{Npt
01 015 reion2step to Convention} 
 compare the
constraints on $r$ by the N-point method and the instantaneous
model. The left and right panels of these plots show the results
for the fiducial $r=0.1$ and $r=0.15$, respectively. The
statistics {describing} them, including best-fit, mean $\bar{r}$,
standard deviation $\sigma$ and CL{,} are shown in Table{s}
\ref{table1} and \ref{table2}. 
  It can be seen that
for a complex reionization history,  the simple instantaneous
assumption generally biases $r$ to a smaller value. In our case,
the true value of $r$ is even ruled out at{ the} 68\% CL for both
double-step reionization and two-step reionization models. The
application of the N-point method can correct for these biases and
the best-fit value of $r$ is constrained to within{ the} 5\% error
level if the true value of $r\gtrsim0.1$. The N-point method also
gives a better inference on $\bar{r}$.

\begin{table}
\centering

\caption{\baselineskip 3.6mm Statistics of 1D marginalized
probability of $r$ for MCMC analysis on the expected
\textit{Planck} pseudo power spectra with double reionization.
\label{table1}}

\vspace{-3mm} \fns \tabcolsep 3mm
\begin{tabular}{l c c c c c}
\hline\noalign{\smallskip}
{Fiducial $r$}& 
{Reionization model}& 
{Best-fit}& 
{$\bar{r}\pm \sigma$}&
{68\% CL}&
{95\% CL}\\
\noalign{\smallskip} \cline{3-6} \hline \noalign{\smallskip}
0.1&Instantaneous&0.048&0.074 $\pm$ 0.043&[0, 0.092]&[0, 0.152]
\\
0.1&N-point&0.095&0.102 $\pm$ 0.045&[0, 0.122]&[0,  0.178]
\\
0.15&Instantaneous&0.143& 0.128 $\pm$ 0.049&[0, 0.149]&[0, 0.210]
\\
0.15&N-point&0.143&0.153 $\pm$ 0.049&[0.128, 0.175]&[0.075, 0.237]
\\
\noalign{\smallskip}\hline
\end{tabular}

\vs \small \centering

\begin{minipage}{120mm}

\caption{\baselineskip 3.6mm Same as Table~\ref{table1},  but for
the expected \textit{Planck} pseudo power spectra with two-step
reionization.\label{table2} }\end{minipage}

\fns \tabcolsep 3mm
\begin{tabular}{ l c c c c c }
\hline\noalign{\smallskip}
{Fiducial $r$}&
{Reionization model}&
{Best-fit}
&
{$\bar{r}\pm \sigma$}&
{68\% CL}&
{95\% CL}\\
\noalign{\smallskip}\cline{3-6}\hline\noalign{\smallskip}
0.1&Instantaneous&0.048& 0.070 $\pm$ 0.039&[0, 0.086]&[0, 0.141]
\\
0.1&N-point&0.047-0.095&0.089 $\pm$ 0.042&[0, 0.107]&[0,  0.164]
\\
0.15&Instantaneous&0.095& 0.117 $\pm$ 0.049&[0, 0.139]&[0, 0.200]
\\
0.15&N-point&0.143&0.140 $\pm$ 0.048&[0, 0.162]&[0, 0.221]
\\
\noalign{\smallskip}\hline
\end{tabular}

\end{table}

\begin{figure}[h]

\vs\centering
\includegraphics[width=3.6cm]{
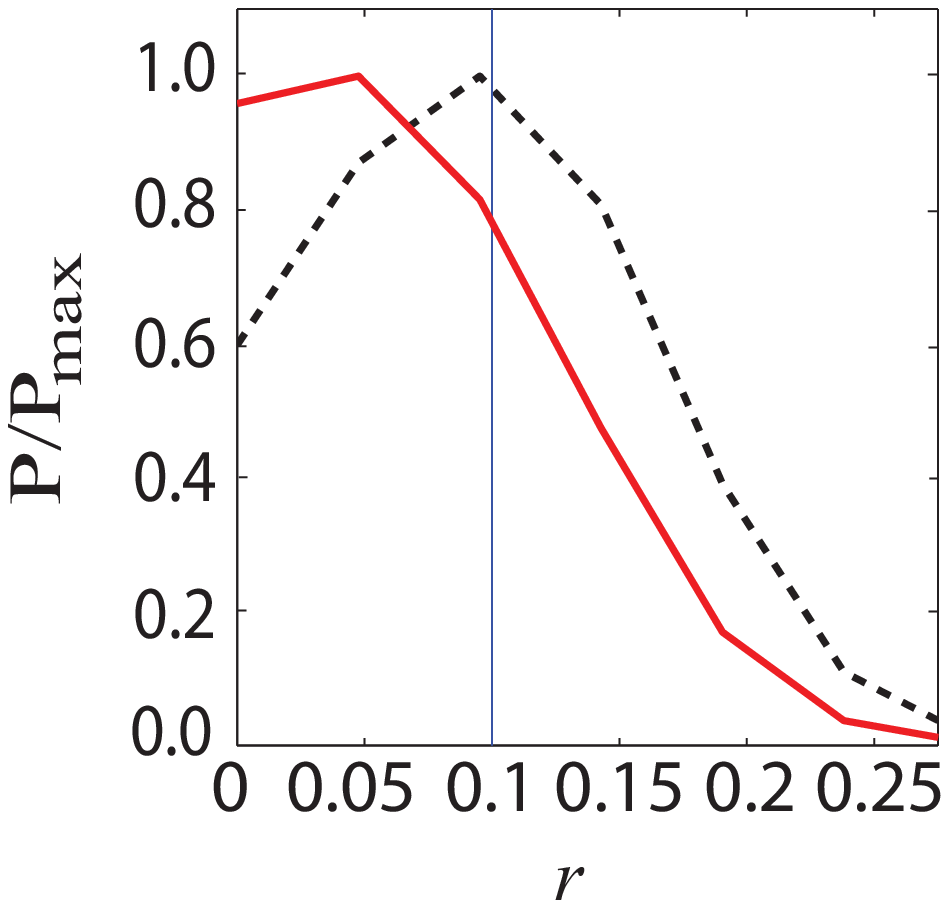}~~~~~
\includegraphics[width=3.6cm]
{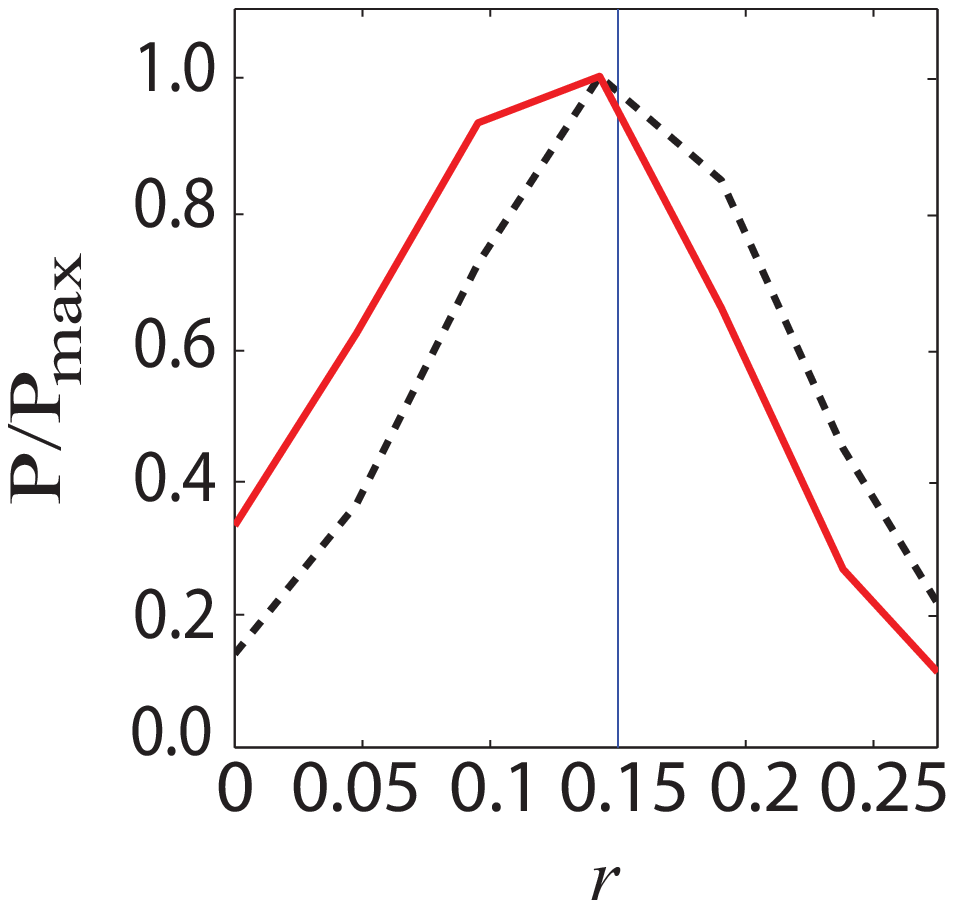}

\vspace{-3mm}
 \caption{\baselineskip 3.6mm
1D marginalized probability $P$ of $r$ (relative to the peak value
$P_{\max}$) in MCMC fitting of the expected \textit{Planck} pseudo
power spectra with double reionization as the fiducial model. The
left and right panels show fiducial $r = 0.1$ and $r = 0.15$ ({\it
indicated by vertical lines}), respectively. The solid and dashed
lines stand for results of the instantaneous and N-point
parameterizations respectively.} \label{Npt 01 015 reion1 to
Convention}
\end{figure}
\begin{figure}

\centering
\includegraphics[width=3.6cm]{
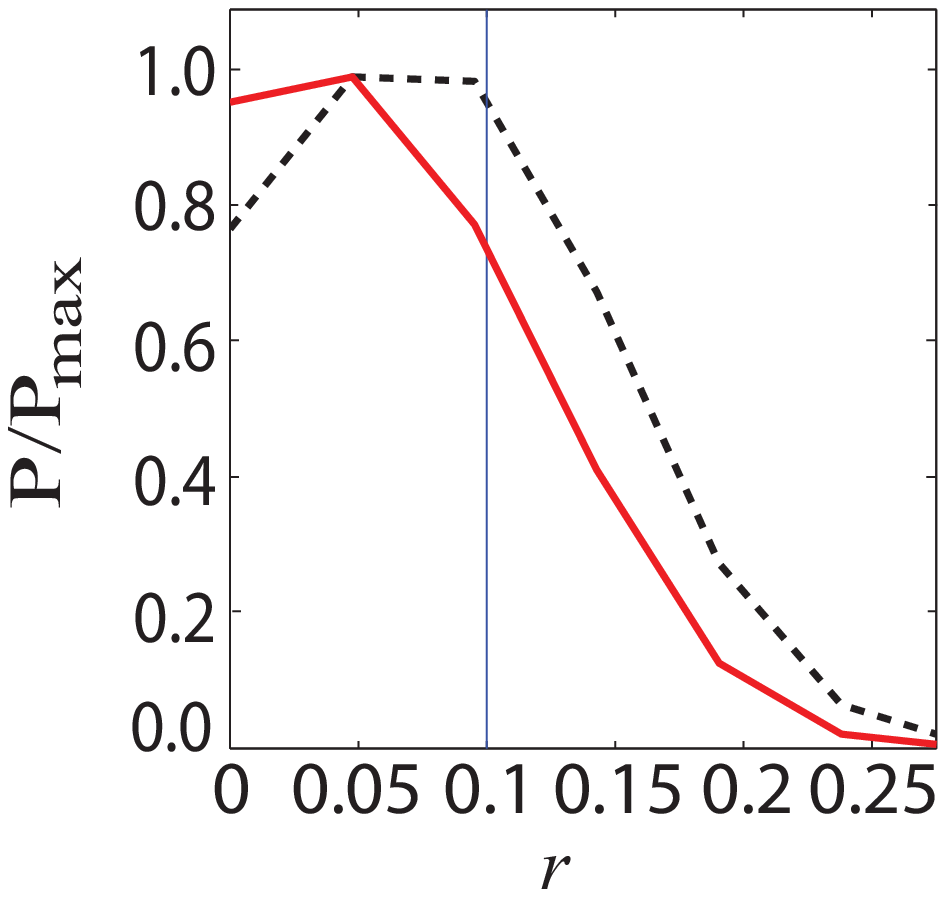}~~~~~
\includegraphics[width=3.6cm]{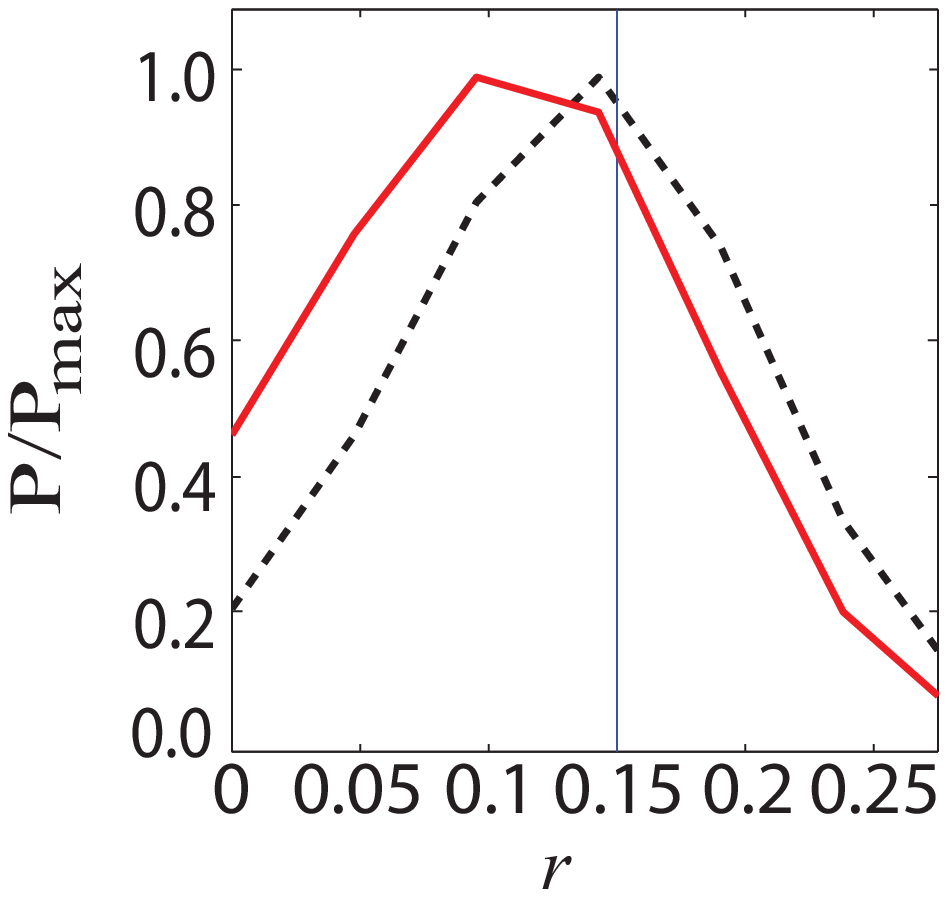}

\begin{minipage}{110mm}
\caption{\baselineskip 3.6mm Same as Fig.~\ref{Npt 01 015 reion1
to Convention}, but with two-step reionization as the fiducial
model.} \label{Npt 01 015 reion2step to Convention}\end{minipage}
\end{figure}

\begin{figure}
\centering
\includegraphics[width=3.6cm]{
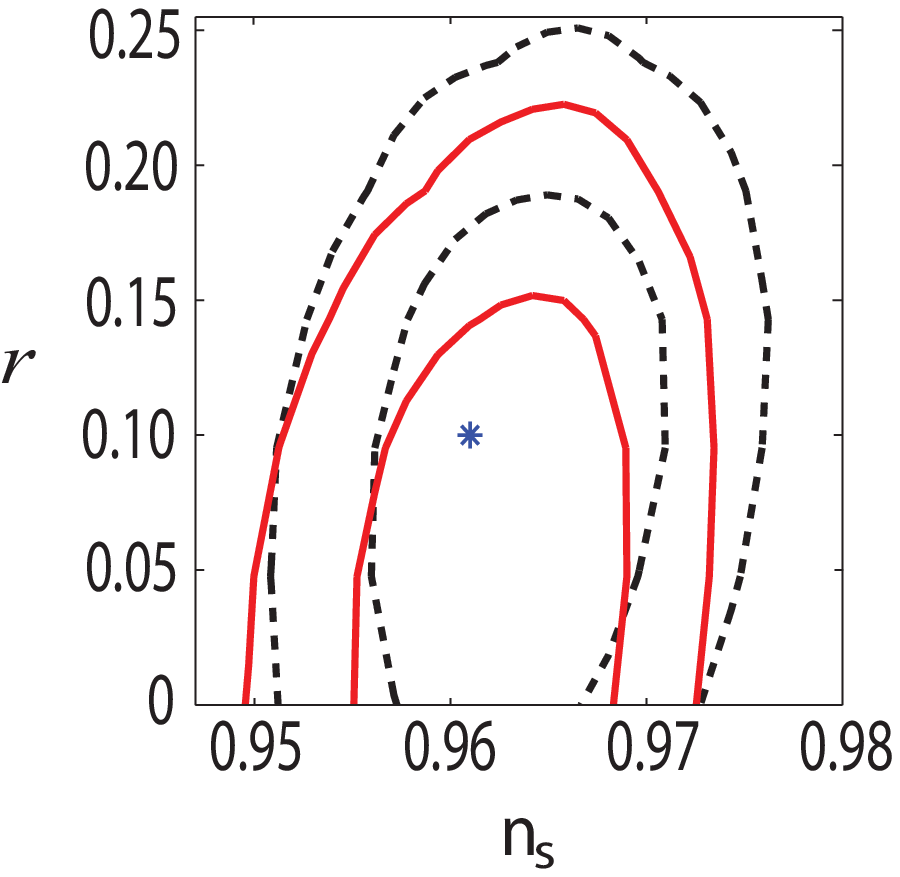}~~~~~~
\includegraphics[width=3.6cm]
{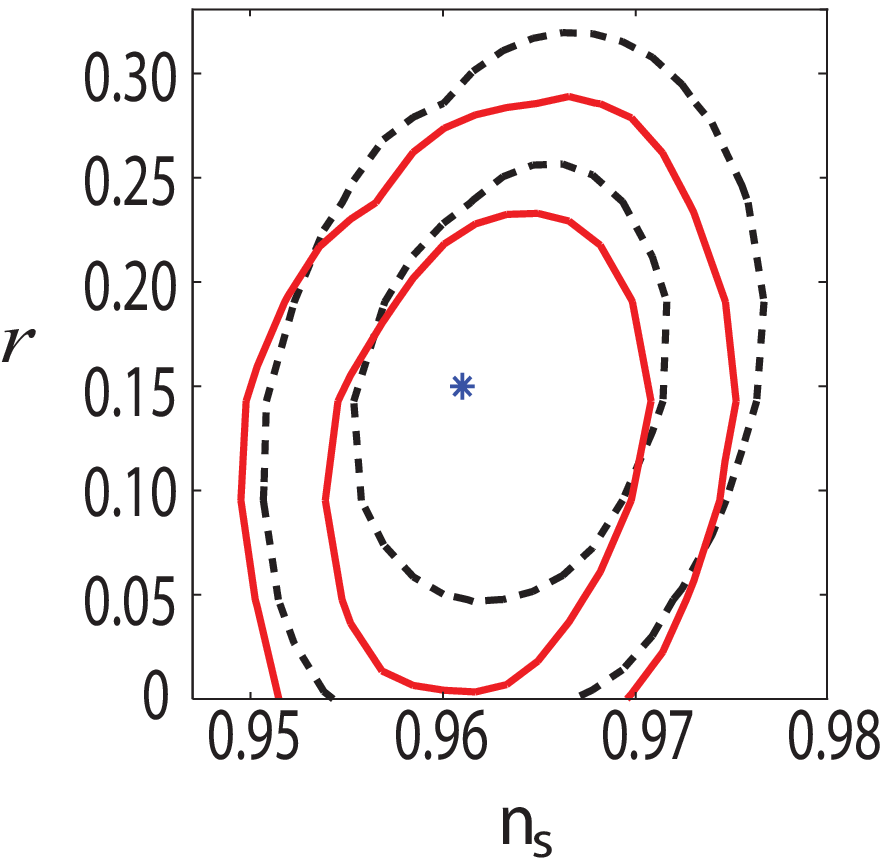}

\vspace{-3mm} \caption{\baselineskip 3.6mm 2D marginalized
probability contours of $r$ vs. $n_{\rm s}$ (68\% CL and 95\% CL)
in{ the} MCMC fitting of the expected \textit{Planck} pseudo power
spectra with double reionization as the fiducial model. The left
and right panels show the results with the fiducial $r = 0.1$ and
$r = 0.15$, respectively.  The solid and dashed lines stand for
results of the instantaneous and N-point parameterizations
respectively. The {asterisks} indicate the fiducial values.}
\label{Npt 01 015 reion1 to Convention 2D}

\vs \centering
\includegraphics[width=3.6cm]{
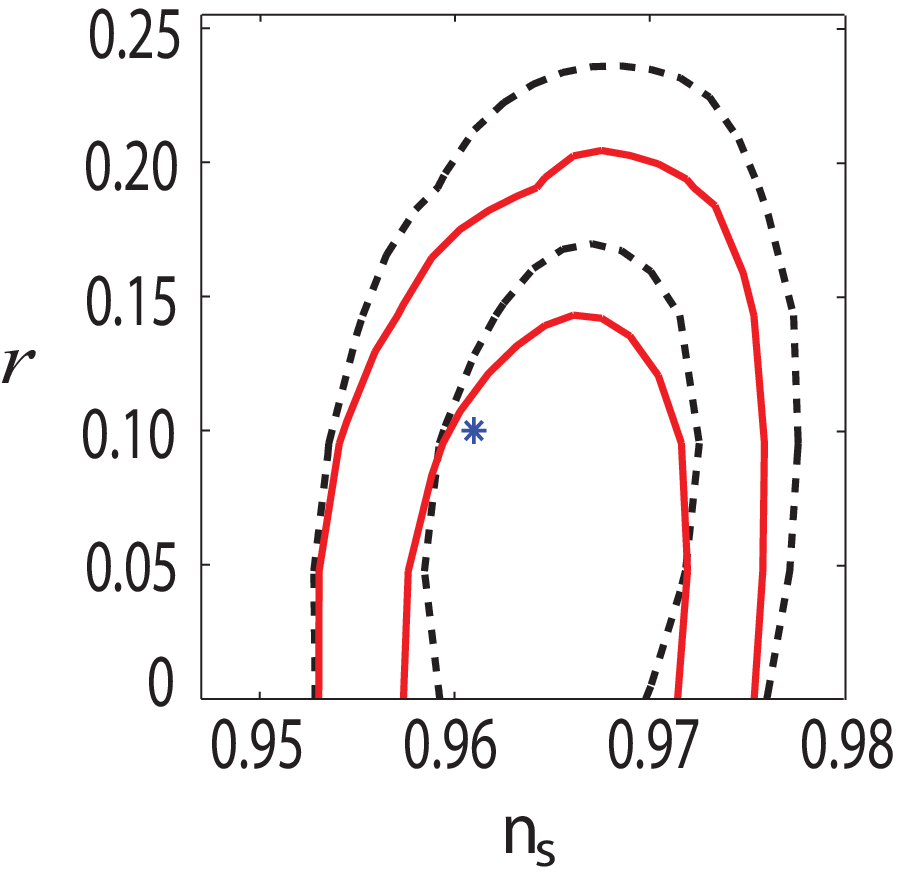}~~~~~
\includegraphics[width=3.6cm]{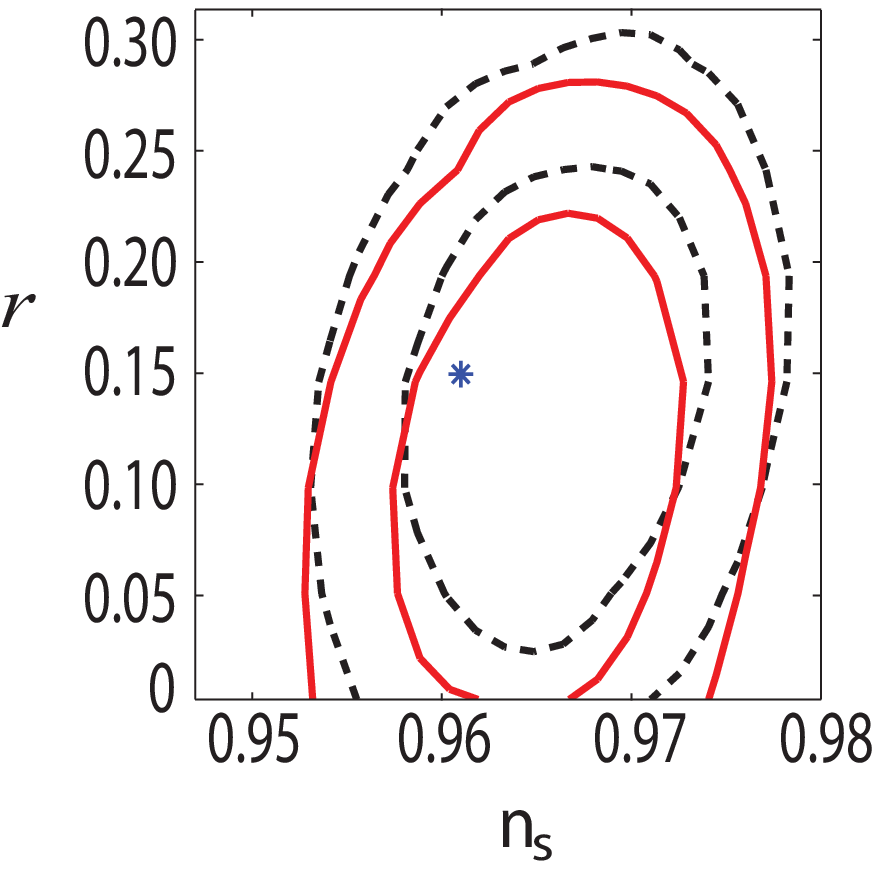}

\begin{minipage}{120mm}
\caption{\baselineskip 3.6mm Same as Fig.~\ref{Npt 01 015 reion1
to Convention 2D}, but with the two-step reionization as the
fiducial model.} \label{Npt 01 015 reion2step to Convention
2D}\end{minipage}

 \centering
\includegraphics[width=5.2cm]{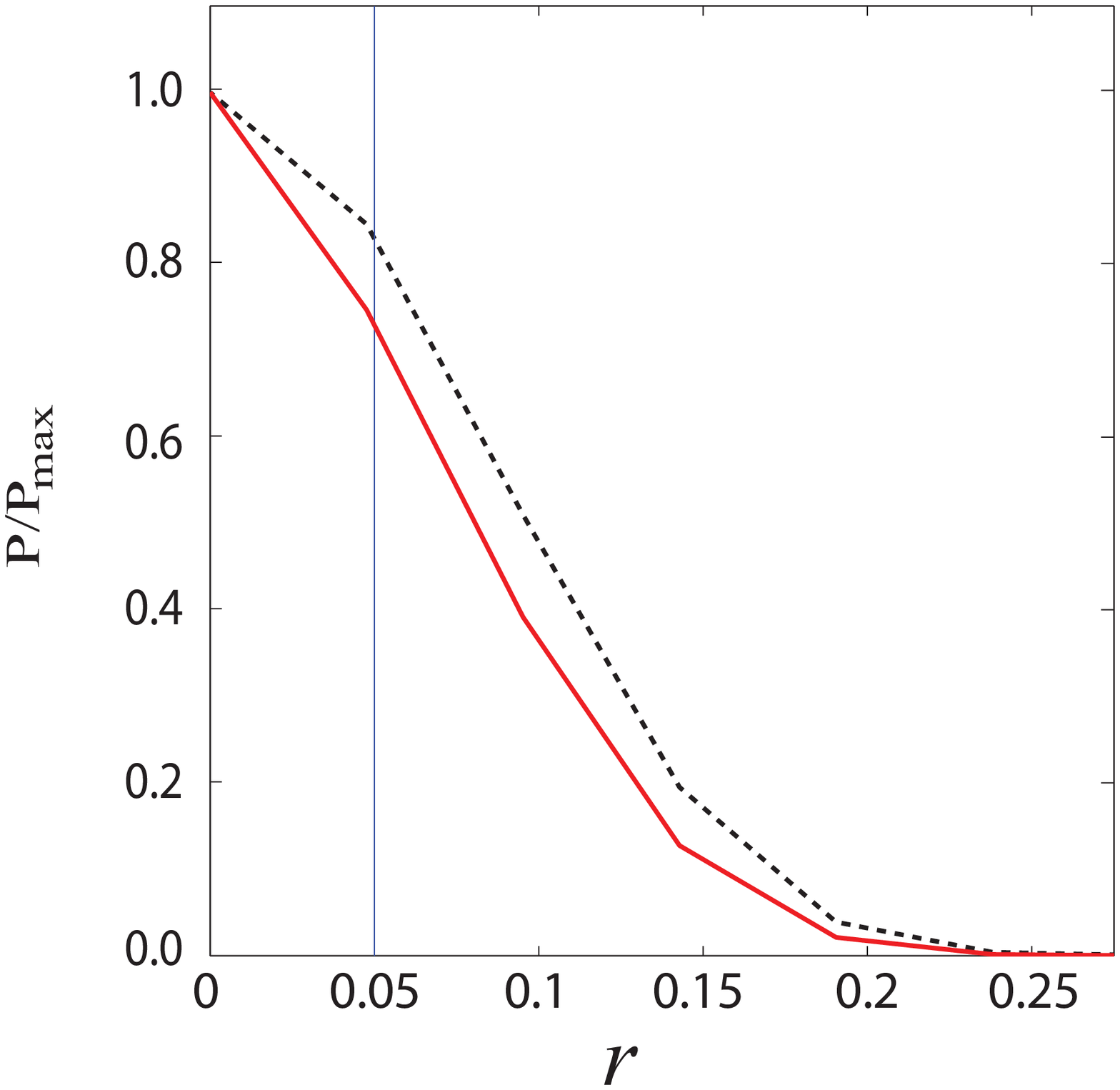}

\vspace{-4mm} \caption{\baselineskip 3.6mm 1D marginalized
probability $P$ of $r$ (relative to the peak value $P_{\max}$)
 in{ the} MCMC
fitting of the expected \textit{Planck} pseudo power spectra with
two-step reionization as the fiducial model, with $r=0.05$ ({\it
indicated by{ a} vertical line}). The solid and dashed lines stand
for results of the instantaneous and N-point parameterizations
respectively.} \label{Npt 005 01 015 reion2step to Convention}
\end{figure}

Figures~\ref{Npt 01 015 reion1 to Convention 2D} 
 and
\ref{Npt 01 015 reion2step to Convention 2D} show the
corresponding 2D contours for $r$ vs. $n_{\rm s}$. In these 2D
cases, the true values of $r$ and $n_{\rm s}$ are not ruled out
at{ the} 68\%{ CL} and 95\% CL for both the N-point method and
instantaneous reionization assumption applied in the analysis.

\begin{figure}

\vspace{-2mm}

\centering
\includegraphics[width=9.cm]{
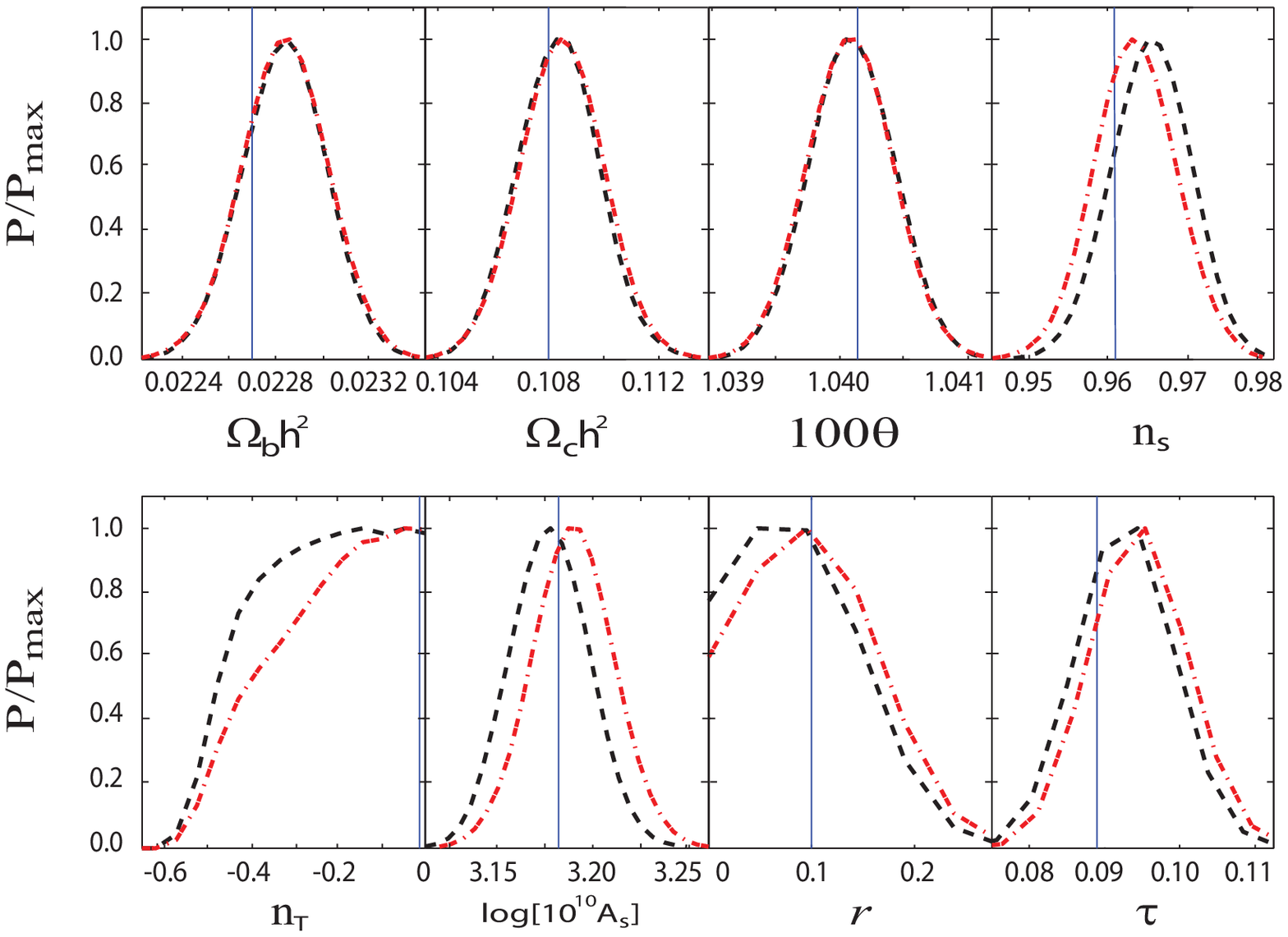}
\includegraphics[width=3.5cm]
{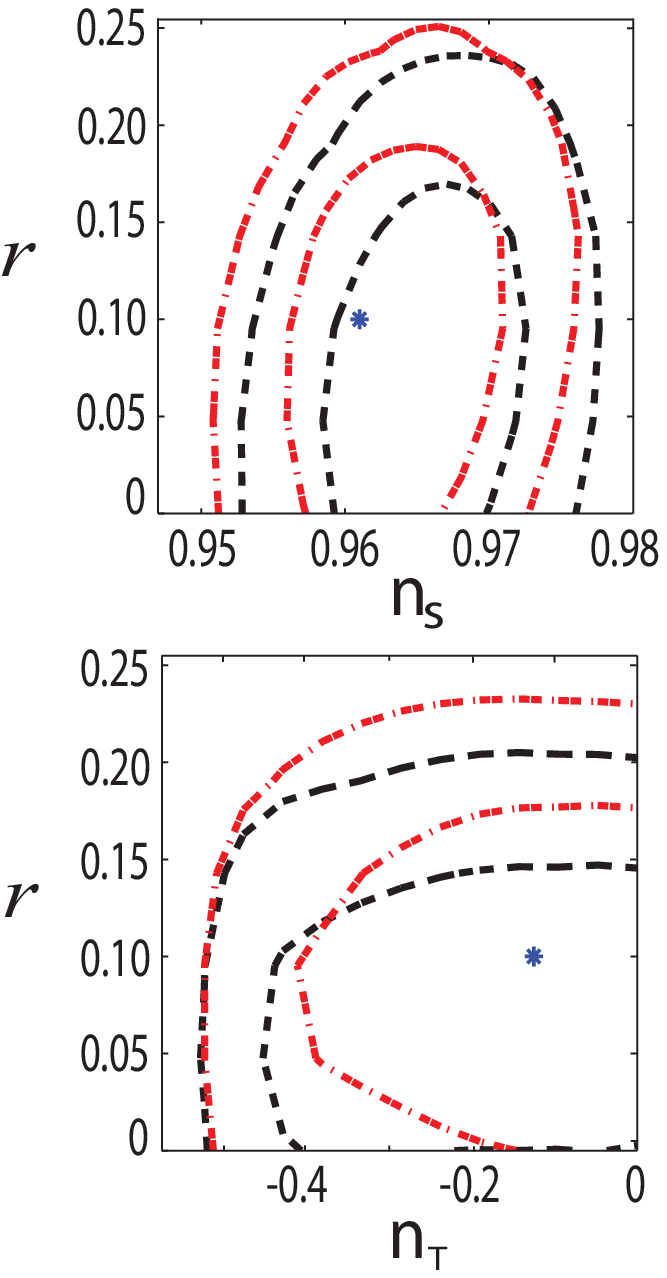}

\vspace{-3mm}

\caption{\baselineskip 3.6mm {\it Left}:
1D marginalized probability $P$ of cosmological parameters
(relative to the peak value $P_{\max}$)  using the N-point method
in {the }MCMC fitting of two sets of the expected \textit{Planck}
pseudo power spectra with fiducial $r$ = 0.1 but with two-step
reionization ({\it dashed lines}) and double reionization ({\it
dashed-dotted lines}). The vertical lines indicate the fiducial
values. {\it Right}: The corresponding 2D marginalized probabilities of
$r$ vs. $n_{\rm s}$ and $r$ vs. $n_{T}$. The 
asterisks 
indicate the fiducial values.}
\label{estimations in cutsky for r=0.1}
\end{figure}

However, the N-point method has its own limitation when $r$ is
small. Figure~\ref{Npt 005 01 015 reion2step to Convention}
 shows the 1D marginalized probability distribution of
$r$ with the fiducial $r=0.05$. It can be seen that the
marginalized probability of the fitted $r$ still prefers $r = 0$,
which indicates that if{ the} N-point method is applied, the
\textit{Planck} data cannot by {themselves} be used to detect $r$
if its value is close to $0$.

\begin{figure}[pth!!]

\centering
\includegraphics[width=9.2cm]{
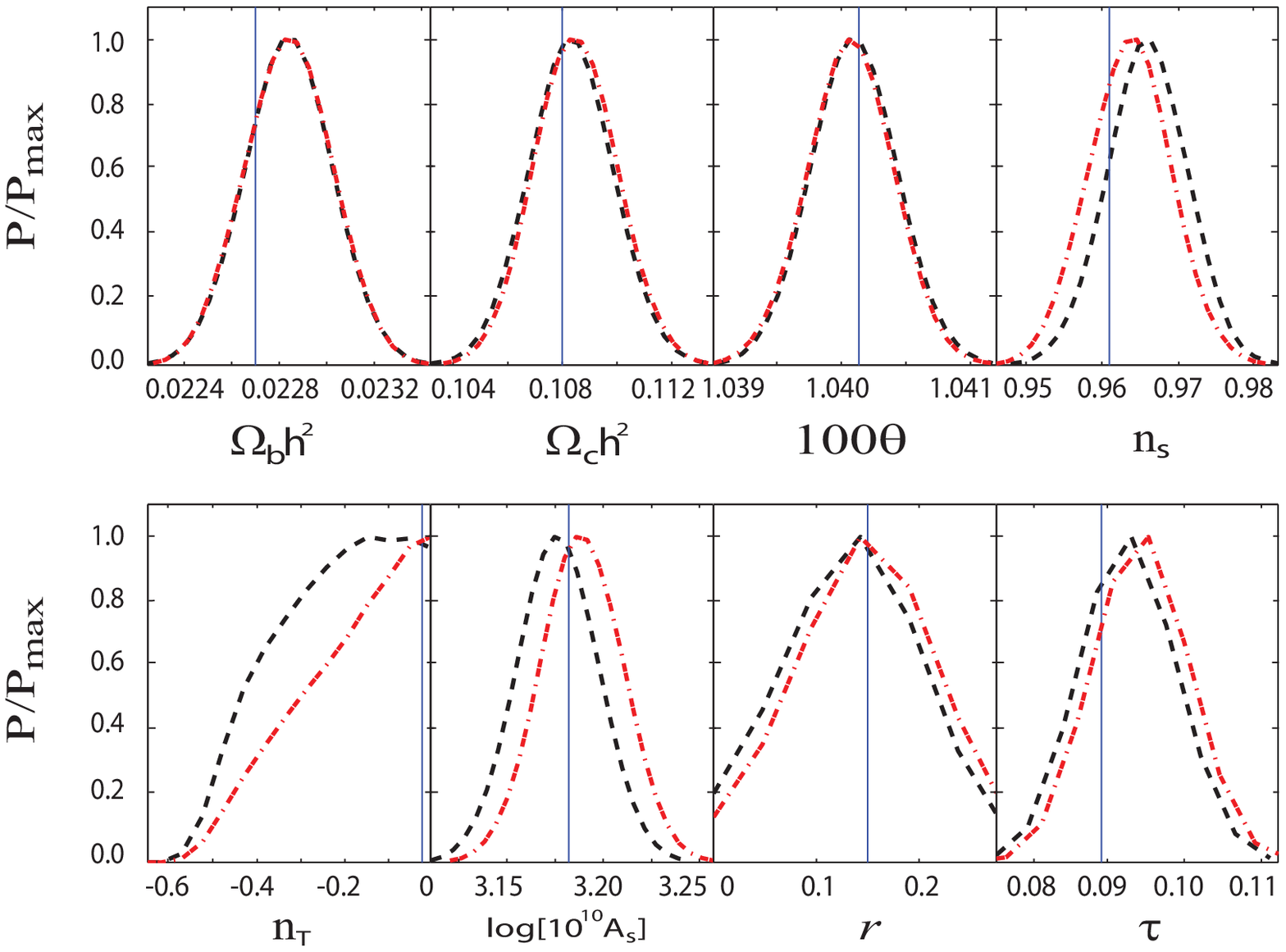}
\includegraphics[width=3.5cm]
{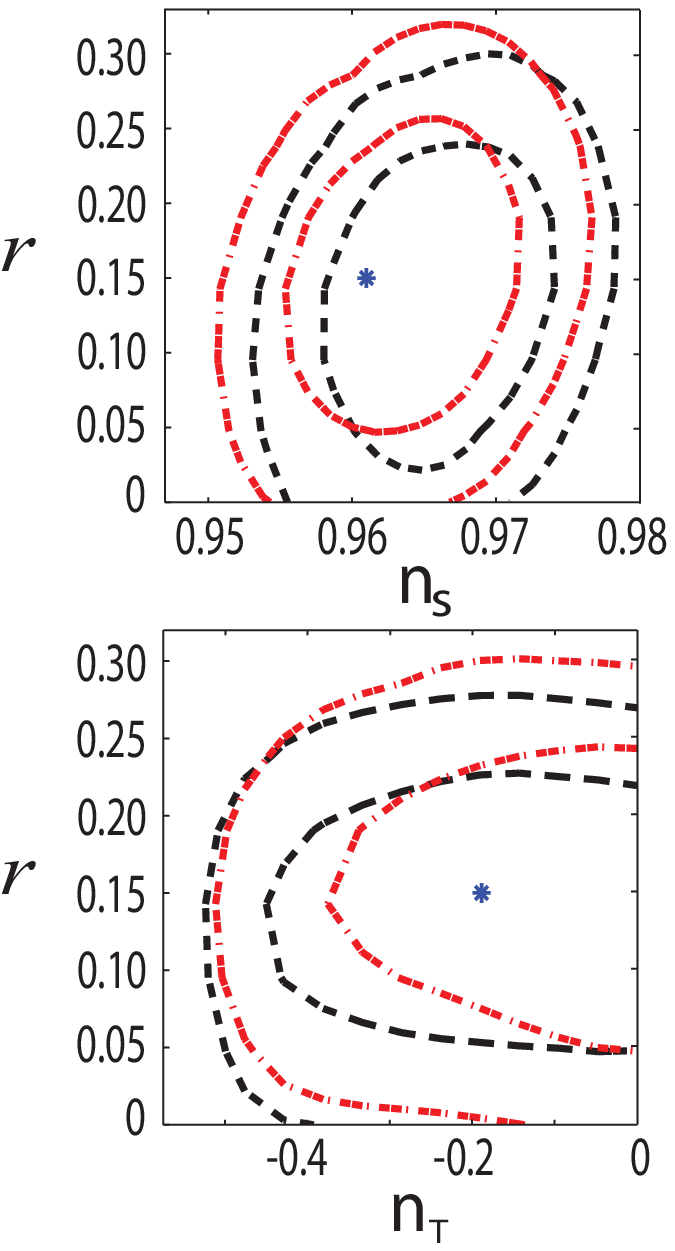}

\begin{minipage}{70mm}
\caption{\baselineskip 3.6mm Same as Fig.~\ref{estimations in
cutsky for r=0.1}, but for $r=0.15$.} \label{estimations in cutsky
for r=0.15}\end{minipage}
\end{figure}

Figures~\ref{estimations in cutsky for r=0.1} 
 and
\ref{estimations in cutsky for r=0.15} show the fitting result of
cosmological parameters applying the N-point method for the
fiducial $r=0.1$ and $0.15$ respectively (we do not repeat the
similar results from $r=0.05$). It shows that the N-point method
with H.L. likelihood also gives reasonable results for the fitting
of other cosmological parameters ($\Omega_{b}h^2, \Omega_{c}h^2, \theta, n_{\rm s}${ and} $\textup{log}(10^{10}A_{\rm s})$). {By} contrast, $\tau$ is still biased with its true value being ruled out at{ the} 68\% CL. {This} is because $\tau$ is sensitive to the MCMC fitting result for the reionization points ($\{ {X_{e}(z_{i})} \}_{i=1}^{N}$ for $N=7$) as it is obtained by {computing an} integration over $X_e(z)$ (Eq.~(\ref{cal_tau})), but our results show poor convergence for them, which means the constraints on the reionization history itself are poor based on 
\textit{Planck}'s power spectra alone. However, the constraint on
$r$ does give satisfactory results in the general reionization
scenario by marginalization since $C_{l}^{\rm BB}$ is not as
sensitive as $\tau$ to the $N$ reionization points, as shown in
Figure~\ref{EE TE BB depends on Xe}.

The 2D contours for $r$ vs. $n_{\rm s}$ and $r$ vs. $n_{\rm T}$
are also shown in the same graphs as they can be used to
discriminate among inflation models and check for the consistency
relation. The true values of $r$, $n_{\rm s}$ and $n_{\rm T}$ are
not ruled out at 68\%{ CL} and 95\% CL for both two-step
reionization and double reionization. These contours also show
that even with \textit{Planck}-quality data, a large degeneracy
still exist{s}
 among $r$, $n_{\rm s}$ and $n_{\rm T}$.

\section{Conclusions}
\label{Conclusion}

It has been pointed out that $r$ can be biased by a{n}
{incorrect} assumption {about} reionization history. In this paper, we consider the general reionization scenario by studying double reionization and two-step reionization models. We apply the {H.L.} likelihood approximation 
to give an idealized constraint of $r$ for the expected \textit{Planck} cut-sky power spectra. We found that the estimation of $r$ is possibly biased by an overly simplistic instantaneous reionization model. The N-point linear interpolation model of reionization can correct this bias if $r\gtrsim0.1$. On the other hand, if $r\lesssim0.1$, not even the N-point method can produce an accurate inference of $r$ from the \textit{Planck} data, given current uncertainties in $X_{e}(z)$.

To calculate the expected \textit{Planck} power spectrum, we
simply add the fiducial power spectrum with the appropriate beam
factor and noise spectrum, following the multiplication of mask
matrices, as shown in Equation~(\ref{expected_spectrum}). We did
not extract the power spectrum from a CMB sky-map generated
randomly according to{ a} Gaussian distribution, as our focus is
on the analysis method applied to account for the reionization
history. Unlike the sky-map approach, which should have $\sim
1\sigma$ scatters from the input in the inference on cosmological
parameters even if the correct cosmic model is applied, we found
that the outcome of $r$ is sharply peaked at its fiducial value up
to $r=0.05$ when the assumption on reionization is correct. Thus,
our approach highlights that deviations shown by the inference on
$r$, if present, are due to the {incorrect} assumption
on cosmic reionization rather than randomness in{ the} sky-map simulation. 

Our conclusion is based on the results using $N=7$ in the N-point
method. In principle, better results may be obtained if $N$ is
further increased so as to improve the modeling of the reionization
history. However, it will also largely increase the convergence time
for the chains in MCMC since the total numbers of varying parameters
are $7+N$. We choose $N=7$ in order to have a reasonable balance
between computation time and good representation of the reionization
history.

In our method, we focus on the cut-sky estimators
$P(b_{l}^{2}C_{l}+N_{l})$ instead of trying to recover the full-sky
CMB power spectra from the cut-sky spectra by imposing the inverse
of the WMAP mask matrices $\{ M^{XY}_{ll^{'}} \}$. This is because
the matrices $\{ M^{XY}_{ll^{'}} \}$ are almost singular and
imposing their inverse on the cut-sky power spectra amplifies the
noise in {them} and thus worsens the forecast. {O}ne of the
advantages of applying the H.L. likelihood {is }that we can change the
CMB full-sky estimators in it to be the cut-sky ones.

We omit the band-power, multiple frequency estimators and the
anisotropic noises here. Moreover, we extend the application of the
H.L. likelihood to $l_{\textup{min}}=2$ for making {the
}forecast, while the likelihood computed from the Internal Linear
Combination maps is usually applied in the range $l \leq 30$
in practice (Dunkley et al.~\citeyear{likelihoodfromILC1}; Larson et
al.~\citeyear{likelihoodfromILC2}). These factors may further limit
the accuracy of the constraints on $r$. We also omit the
contamination of the B-mode power spectrum by{ the} gravitational lensing
effect as it is expected to be removed (Seljak \&
Hirata~\citeyear{lensingonBBbyseljak}). In the future, we would like
to explore this problem using a more realistic method, for instance,
using the full likelihood in the low-$l$ range and model the
\textit{Planck}-quality data from{ a} simulated CMB sky-map.

\begin{acknowledgements}
This work is partially supported by a grant from the Research
Grant Council of the Hong Kong Special Administrative Region,
China (Project No. 400910). J. TANG {is grateful for} the support
of a postdoctoral fellowship by The Chinese University of Hong
Kong. We also thank the ITSC of The Chinese University of Hong
Kong for providing clusters{ that were used} for computations.
\end{acknowledgements}

\appendix
\section{Hamimeche-Lewis likelihood}
\label{appendix}

The H.L. likelihood approximation gives the maximum likelihood
function (Hamimeche \& Lewis~\citeyear{Lewis2008}) as
\begin{equation}
-2\textup{ln}\mathcal{L}=\mathbf{X}_{g}^{T}\mathbf{M}_{f}^{-1}\mathbf{X}_{g} 
=\sum_{ll^{'}}[\mathbf{X}_{g}]^{T}_{l}[\mathbf{M}_{f}^{-1}]_{ll^{'}}[\mathbf{X}_{g}]_{l^{'}}\label{HL},
\end{equation}
where $\mathbf{M}_{f}$ is the covariance block matrix for $\mathbf{X}_{l}=\textup{vecp}(\mathbf{C}_{l})$ (a vector of all distinct elements of matrix $\mathbf{C}_{l}$) and {is }evaluated for some specific fiducial model $\{{C}_{l}\}=\{{C}_{fl}\}$. It contains $(l_{\textup{max}}-l_{\textup{min}}+1) \times (l_{\textup{max}}- l_{\textup{min}}+1)$ blocks labeled by $l$ and $l^{'}$, and $\mathbf{X}_{g}$ is generally a $(l_{\textup{max}}-l_{\textup{min}}+1)\times6$ row block vector:
\begin{equation}
[\mathbf{M}_{f}]_{ll^{'}}=\langle(\hat{\mathbf{X}}_{l}-\mathbf{X}_{l})(\hat{\mathbf{X}}_{l^{'}}-\mathbf{X}_{l^{'}})^{T}\rangle_{f}
\end{equation}

\begin{equation}
[\mathbf{X}_{g}]_{l}= \textup{vecp}\left(\mathbf{C}_{fl}^{1/2}g[\mathbf{C}_{l}^{-1/2}\mathbf{\hat{C}}_{l}\mathbf{C}_{l}^{-1/2}]\mathbf{C}_{fl}^{1/2}\right)
\end{equation}
\begin{equation}
\mathbf{C}_{l}=
\begin{pmatrix}
C_{l}^{\rm TT} & C_{l}^{\rm TE} & C_{l}^{\rm TB} \\[0.3em]
C_{l}^{\rm TE} & C_{l}^{\rm EE} & C_{l}^{\rm EB} \\[0.3em]
C_{l}^{\rm TB} & C_{l}^{\rm EB} & C_{l}^{\rm BB}
\end{pmatrix}
\end{equation}
\begin{equation}
\mathbf{C}_{fl}=
\begin{pmatrix}
{C_{fl}^{\rm TT}} & {C_{fl}^{\rm TE}} & {C_{fl}^{\rm TB}} \\[0.3em]
{C_{fl}^{\rm TE}} & {C_{fl}^{\rm EE}} & {C_{fl}^{\rm EB}} \\[0.3em]
{C_{fl}^{\rm TB}} & {C_{fl}^{\rm EB}} & {C_{fl}^{\rm BB}}
\end{pmatrix}
\end{equation}
for $l, l^{'}=l_{\textup{min}}, \cdots, l_{\textup{max}}$, where
\begin{equation}
g(x)\equiv \textup{sign}(x-1)\sqrt{2(x-\textup{ln}x-1)}
\end{equation}
\begin{equation}
[g(\mathbf{A})]_{ij} =
\begin{cases}

g(\mathbf{A}_{ii})\delta_{ij} & \mathbf{A} \; \text{is diagonal}\\

[\mathbf{U}g(\mathbf{D})\mathbf{U}^{\rm T}]_{ij} & \mathbf{A} \;
\text{is symmetric positive-definite}
\end{cases}
\end{equation}
(then $\mathbf{A}=\mathbf{U}\mathbf{D}\mathbf{U}^{\rm T}$ for some diagonal
 matrix $\mathbf{D}$). The assumption $C_{l}^{\rm TB}=C_{l}^{\rm EB}=0$
  is applied in our study.
Equation (\ref{HL}) gives the exact results for the full sky
$C_{l}$.  Moreover, it has been tested to be reliable {in the}
range $30<l<2000$ when used on the masked-sky spectra $P(C_{l})$.
To deal with the cut-sky effect, all of the full-sky power spectra
$C_{l}$ described above are changed{ such that}
\begin{equation}
C_{l} \rightarrow P(b_{l}^{2}C_{l}+N_{l}).
\end{equation}

For the computation of the covariance matrix $\mathbf{M}_{f}$, the
fiducial model we applied is based on the same cosmological
parameters as the input, since they are well constrained by present
cosmological surveys, but we fixed $r=0.15$ and used the
instantaneous reionization model with $\tau=0.089$ and
$z_{re}=10.5$. Using HEALPix\footnote{HEALPix is available at {\it
http://healpix.sourceforge.net/}} (G\'{o}rski et
al.~\citeyear{healpix1}), we can compute $\mathbf{M}_{f}$ by
generating random samples from the same $C_{l}$ power spectra. As
there are $p=6 \times (l_{\textup{max}}-l_{\textup{min}}+1) $
estimators in $\{\mathbf{X}_{l}\}_{l}$, we generate $1.5 \times
10^{5}$ random samples in order to have a good convergence,
following the $N_{\textup{sample}} \sim p\textup{ln}p$ rule
(Vershynin~\citeyear{Vershynin}).

\end{document}